\documentclass{aa}  

\usepackage{array}

\usepackage{graphicx}

\usepackage{txfonts}

\usepackage{natbib}
\usepackage{hyperref}

\begin{document}

   \title{Magnesium isotopic detection in cool stars: Tracing nucleosynthetic signatures from MgH features}
    
   \titlerunning{Magnesium Isotopes in Cool Stars}

   \author{Q.~Aicken Davies
          \inst{1}
          \and C.~C.~Worley \inst{1}}

   \institute{$^{1}$University of Canterbury, Private Bag 4800, Christchurch 8140, New Zealand \\
   \email{quinastronomy@gmail.com}}

   \date{Received}

  \abstract
    {Magnesium isotopic ratios offer valuable insights into stellar nucleosynthesis and Galactic chemical evolution, particularly in distinguishing between contributions from supernovae and asymptotic giant branch (AGB) stars. These isotopes are accessible through MgH molecular features in cool stellar atmospheres, yet their measurement remains challenging across a range of spectral types.} 
   {We aim to assess the reliability of MgH spectral regions for extracting magnesium isotopic ratios ($^{24}$Mg, $^{25}$Mg, and $^{26}$Mg) in stars spanning spectral types from M to G, and to evaluate the consistency of these measurements with nucleosynthetic expectations.}
   {We applied an analysis pipeline using spectrum synthesis to derive isotopic ratios, validated using three well-studied reference stars, to a sample of five additional dwarf and giant stars. Individual MgH molecular band regions were analysed to determine their sensitivity to isotopic variation. Europium (Eu) and barium (Ba) abundances were also measured to explore potential correlations with magnesium isotopic ratios as $r$ and $s$~process proxies, respectively.}
   {There are ten wavelength regions for which MgH have previously been investigated. Our study determined that seven of these regions were the most reliable for extracting isotopic information. Other regions exhibited limited sensitivity between stellar type and parameters. The Mg isotope ratios ($^{24}$Mg:$^{25}$Mg:$^{26}$Mg) obtained in this work include: HD~11695-81:7:12; HD~18884-81:7:12; HD~18907-69:9:23; HD~22049-71:16:13; HD~23249-66:13:22; HD~128621-67:17:16; HD~10700-78:10:12; and HD~100407-65:10:25. Comparison of europium (Eu) abundances with the three magnesium isotopes revealed strong correlations. The strongest correlation was with $^{24}$Mg. Note that $^{24}$Mg  is predominantly produced by hydrostatic $\alpha$ capture in massive stars, a process that precedes the $r$ process responsible for Eu production. In contrast, barium (Ba) abundances showed no significant correlation with $^{25}$Mg and $^{26}$Mg, despite their shared AGB origin.With removal of the effect of metallicity similar results were found.}
   {Our results demonstrate that selected MgH regions can reliably measure magnesium isotopes in cool stars, providing a reproducible framework for future studies of stellar nucleosynthesis and galactic chemical evolution.}

    
   \maketitle

\section{Introduction}\label{sec:intro}
    The nucleosynthetic processes that occur within the later stages of stellar evolution are important for understanding element evolution in the Milky Way. During these processes various elements, and therefore various isotopes of each element, are created in multiple nucleosynthetic environments. Previous studies, \cite{karakas_mg_2003}, \cite{iliadis_nuclear_2007}, \cite{vangioni_cosmic_2019} and \cite{kobayashi_origin_2020}, describe the nucleosynthetic origins of magnesium (Mg) isotopes, amongst other isotopes. There are four nucleosynthetic processes that we subsequently focus on: the $\alpha$ process; hot-bottom burning;  the $s$ process (slow neutron capture); and the $r$ process (rapid neutron capture). 
   
   The $\alpha$ process is one of the main production channels for the $^{24}$Mg isotope. The hydrostatic $\alpha$ process occurs during the Carbon-, Neon-, and Oxygen-core-burning stages of massive stars ($\leq8M_{\odot}$) and generates elements with an atomic number of multiples of four \citep{vangioni_cosmic_2019,iliadis_nuclear_2007}. Such massive stars often end their life as core-collapse supernovae, which is when this isotope is dispersed through the inter-stellar medium (ISM). 

   Hot-bottom burning occurs in the asymptotic giant branch (AGB) phase of intermediate mass stars (0.8-8 $M_{\odot}$). At this stage the stars are hot enough for efficient proton capture by Mg leading to Al which then decays into heavier Mg isotopes. The $^{25,26}$Mg isotopes are also produced during thermal pulses of helium shell burning. Note that $\alpha$ capture on $^{22}$Ne can lead to both $^{25}$Mg and $^{26}$Mg \citep{vangioni_cosmic_2019} (in the $^{22}Ne(\alpha,n)^{25}Mg$ and $^{22}Ne(\alpha,\gamma)^{26}Mg$ reactions \citep{iliadis_nuclear_2007}).

    We note that $s$~process nucleosynthesis occurs during the AGB phase in stars of mass range 0.5 to 8 M$_{\odot}$. The $^{25}$Mg and $^{26}$Mg isotopes are mostly generated via hot-bottom burning, the products are then distributed into the inter-stellar medium (ISM), along with $s$~process elements, via stellar winds \citep{vangioni_cosmic_2019, iliadis_nuclear_2007, karakas_mg_2003}.
    
    Supernovae and high-energy collisions, such as neutron star mergers, are where the $r$~process nucleosynthesis occurs. Small amounts of $^{24}$Mg, $^{25}$Mg, and $^{26}$Mg can be generated by $r$~process nucleosynthesis \citep{kobayashi_origin_2020}.

   Stars that we observe today formed from polluted material in the ISM, and bear the signatures of these different nucleosynthetic events. The isotopes of magnesium can be generated in multiple nucleosynthetic environments, determining the relative isotopic abundances allow us to distinguish between the nucleosynthetic channels that created the elements throughout time. Using isotopes to distinguish between these events allows for a deeper understanding of the history of the galaxy and its chemical evolution.

    Magnesium isotope ratios can be measured from the magnesium hydride (MgH) molecular features present in stellar spectra. In this study, we examined a set of MgH features in order to derive the ratios of $^{24}$Mg, $^{25}$Mg, and $^{26}$Mg. The MgH molecules thermally dissociate due to the covalent bonds that hold the molecules together, which decay at high temperatures \citep{pavlenko_electronic_2008}. Using the Saha equation and a dissociation energy from \citet{shayesteh_ground_2007}, we calculated an effective temperature of $T_{\mathrm{eff}}\approx5800 \pm 300K$, which corresponds to an expected dissociation temperature. This provides a potential upper limit on the stellar effective temperature above which the MgH feature is unlikely to be detectable.

   The Gaia benchmark stars (GBS) are well-studied, bright, or nearby stars of the late F, G, and K spectral types with a range of luminosities and metallicities \citep{heiter_gaia_2015}, making them ideal for examining MgH over a broad range of stellar parameters. The GBS dataset is well know in the galactic archaeology literature and has been widely used in spectroscopic surveys such as GALAH \citep{buder_galah_2021} and Gaia-ESO \citep{gilmore_gaia_2022}. The GBS were curated to be used to calibrate spectroscopic pipelines and make comparison of stellar parameters between spectroscopic surveys. The original set of GBS (V1/V2) contained 34 stars ranging in effective temperature from $4000~K$ to $6500~K$ which included FGKM giants, dwarfs, and subgiants. Given their well-characterised nature and broad coverage of stellar parameters, the GBS provided an ideal sample for determining Mg isotopes.

    Previous studies of Mg isotope determination include \cite{mcwilliam_isotopic_1985}, \cite{barbuy_magnesium_1985}, \cite{barbuy_magnesium_1987}, \cite{mcwilliam_isotopic_1988}, \cite{gay_isotopic_2000}, \cite{yong_magnesium_2003, yong_magnesium_2004, yong_mg_2006}, \cite{melendez_rise_2009}, \cite{thygesen_chemical_2016}, \cite{mckenzie_magnesium_2024}, and \cite{mckenzie_complex_2024}. These works used a range of computational approaches, including differential analysis, manual isotope fitting, and $\chi^2$ fitting to compare synthetic spectra with observations (typically generated with MOOG; \citealt{sneden_moog_2012}). Many of these studies focused on giant stars with a few focusing on cool dwarfs \citep{yong_magnesium_2003,mckenzie_magnesium_2024}. This study builds on these foundations, drawing on their insights into line list development, synthesis, and fitting techniques. To extend this work across a broader range of stellar parameters, we focus on the Gaia benchmark stars. All of these studies have focused on 1D determinations of Mg isotopes. However, several works have also investigated the 3D effects of MgH line formation and how 3D line asymmetries influence the inferred isotopic ratios. These studies aim to quantify how Mg isotope determinations in 3D differ from those obtained in 1D. In general, the impact on the isotopic ratios is small, but in some cases the differences become significant. When this occurs, 1D analyses can misestimate the isotopic ratios by up to 5 percentage points for $^{24}$Mg and $^{25}$Mg \citep{thygesen_investigation_2017}. While this distinction is not critical for the present study, it is worth noting.

    This paper is structured as follows. Section~\ref{sec:observations} describes the observations and data processing. Section~\ref{sec:3} details the measurement of magnesium isotopes and the underlying calculations. In Section~\ref{sec:discussion}, we assess the quality of the data produced by our pipeline, compare our results with literature values, and explore correlations between magnesium isotope abundances and neutron-capture elements. In Section~\ref{sec:conclusion} we engage in final discussions and conclusions.

\section{Observations}\label{sec:observations}

Building on the strengths of the GBS, we selected a subset of stars observable from the University of Canterbury Ōtehīwai Mt John Observatory to investigate MgH features in detail. We used the High Efficiency and Resolution Canterbury University Large Echelle Spectrograph ($\textsc{HERCULES}$) instrument in this study. $\textsc{HERCULES}$ is fibre-fed from the 1~m McLellan telescope \citep{hearnshaw_Hercules_2002,hearnshaw_mt-john_1986}. We chose to use $\textsc{HERCULES}$ to test the capabilities of the instrument with an $R\sim82{,}000$ to confirm if it was capable of isotope determination. The spectra were reduced using the in-house reduction pipelines $\textsc{Megara}$ (Version 1.5) \footnote{\url{https://github.com/DrEmstar/MEGARA}} and $\textsc{Deianira}$ \footnote{\url{https://github.com/qai11/Deianira}}. 

\begin{figure}
    \centering
    \includegraphics[width=0.45\textwidth]{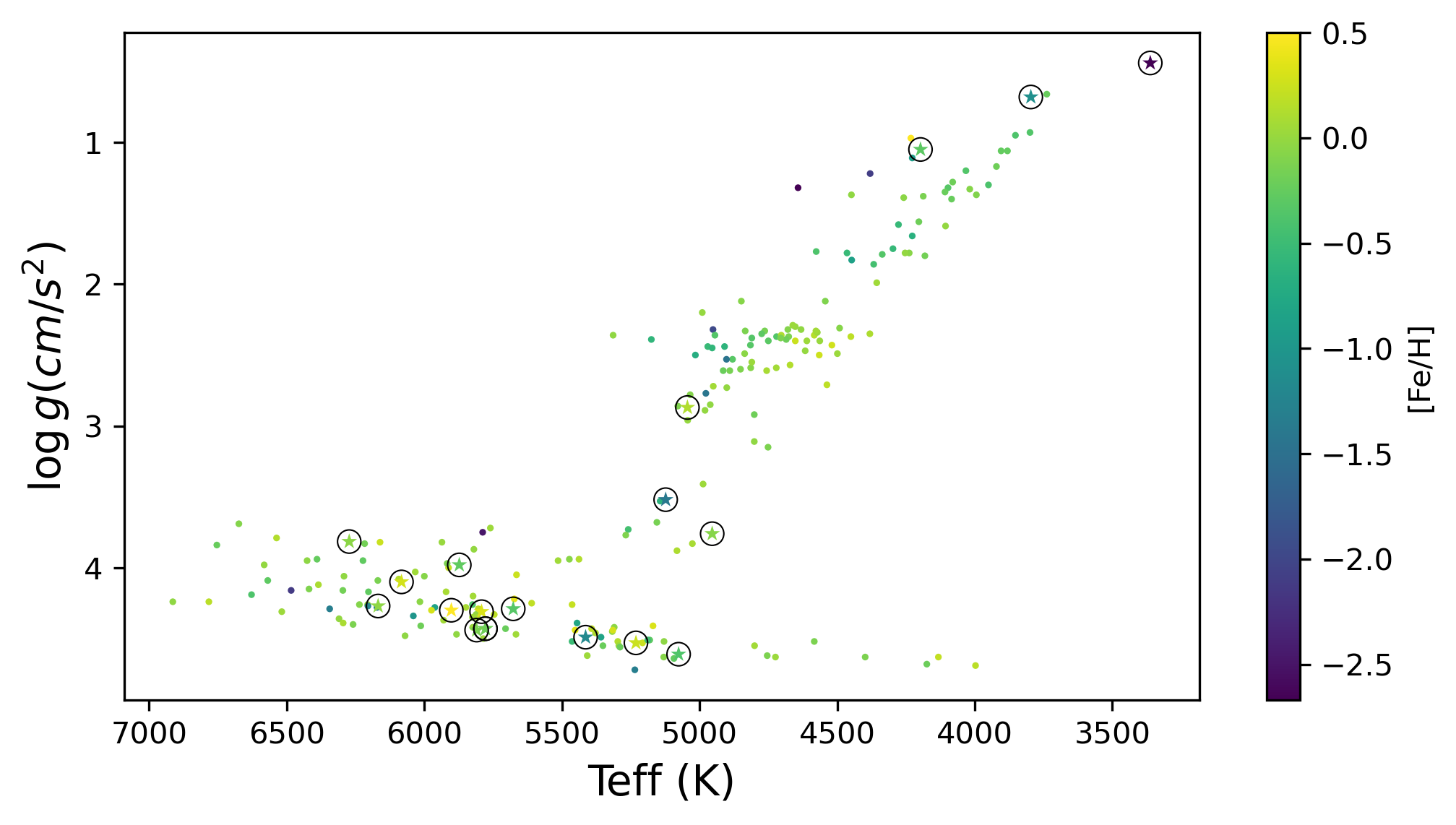}
    \caption{Kiel diagram showing the distribution of ISAM stars in temperature space (circled), coloured by metallicity. The background points are the stars from GBS V3 \citep{soubiran_gaia_2024} to show the main sequence more clearly.}
    \label{fig:Kiel_diagram}
    \vspace{-0.5cm}
\end{figure}

Three observing runs of seven days, were conducted in February, June, and September of 2024. Figure \ref{fig:Kiel_diagram} presents the GBS V3 catalogue \citep{soubiran_gaia_2024} as a Kiel diagram, with the stars observed in this study indicated with circles.

For this study 18 stars were observed, the details of which are shown in Table \ref{tab:results_params}. In order to find Mg isotopes across the parameter space 16 of 18 stars from the GBS were selected that have spectral types ranging from M4 to F9 with both dwarf and giant stars included. Three other stars of F and G types were selected from the 5$^{th}$ catalogue of nearby stars \citep{golovin_fifth_2023} as filler stars that were observable during observing runs. All stars in the Isotope Sample (ISAM) were brighter than V = 6, ensuring the highest signal-to-noise ratio achievable with a 1~m telescope. Between ten and 40 observations were taken for each star, combining observations with a median stack to further increase the signal-to-noise ratio, aiming for a final S/N of approximately 100. We use the modules developed in iSpec \citep{blanco-cuaresma_ispec_2015} to perform the median-stacking, thereby maximising the likelihood of successfully extracting isotope information.
\vspace{-0.5cm}
\subsection{Data processing}

This section outlines the derivation of stellar parameters used to assess continuum normalisation in regions sensitive to MgH. We used the GBS V2 parameters \citep{jofre_gaia_2015} to create our synthetic spectra, as they were derived using robust methods; however, careful continuum normalisation was still required to ensure an accurate comparison with observed spectra. We independently derived stellar parameters using the spectral synthesis tools iSpec \citep{blanco-cuaresma_determining_2014} and MOOG \citep{sneden_moog_2012}. The derived parameters were then compared to the GBS (V2) values, to check for consistency and confirm that the normalisation was valid. The parameters we derived include, effective temperature ($T_{\mathrm{eff}}$), surface gravity ($\log g$), metallicity ([M/H]), projected rotational velocity ($v \sin i$), and $\alpha$-element abundance ([$\alpha$/Fe]). These were iteratively adjusted within iSpec by fitting synthetic spectra to the observed data, allowing convergence on the best-fit values. This approach ensured that our analysis was aligned with established GBS parameters and that the spectra were prepared for isotope extraction. 

We used the same methods as \citet{jofre_gaia_2015} to derive parameters, as described above, a comparison between the derived parameters and the GBS V2 values is shown in Table~\ref{tab:results_params}. For dwarf stars ($\log g > 3.2$), there was good agreement, with all parameters falling within $3\sigma$ of the GBS values. For giant stars ($\log g < 3.2$), the agreement was poorer, with discrepancies up to $5\sigma$ across all parameters. The disagreement between values was expected to some extent. Additionally, the parameters for HD~11695 and HD~157244 highlight a known limitation in iSpec \citep{blanco-cuaresma_determining_2014} where convergence of parameters, particularly $\log g$, can vary significantly. The cause of this limitation was not identified in this study or by \citet{blanco-cuaresma_determining_2014}. However, since the majority of the sample lies within $3\sigma$ of the reference values for most parameters, we conclude that our continuum placement is sufficiently accurate to derive magnesium isotope abundances with reasonable confidence.

\begin{table*}[!htb]
      \caption{Derived stellar parameters for the ISAM.}
         \label{tab:results_params}
         \centering
         \begin{tabular}{ccccccccc}
            ID1 & ID2  &SPT& $\it{T_{\mathrm{eff}}}$ & $\it{T_{\rm{eff},\rm{ref}}}$& $\log$ g & $\log{g}_{\rm{ref}}$ & [M/H]  & [M/H]$_{\rm{ref}}$\\ 
            & && $(K)$ &$(K)$ &$(cm/s^{-2})$&$(cm/s^{-2})$& $dex$ & $dex$\\
            \hline
            \hline
             $\psi$~Phe &HD 11695 &M4III& 3710 $\pm$ 60 & 3472 $\pm$ 92& 1.62 $\pm$ 0.17 & 0.51 $\pm$ 0.18& -0.42 $\pm$ 0.06 &–1.24 $\pm$ 0.39\\  
             $\alpha$ Cet &HD~18884 &M1.5IIIa& 3841 $\pm$ 68 & 3796 $\pm$ 65& 1.59 $\pm$ 0.22 & 0.68 $\pm$ 0.29& -0.16 $\pm$ 0.08 &–0.45 $\pm$ 0.47\\  \hline 
             $\beta$ Ara &HD~157244 &K3Ib-II& 4363 $\pm$ 7 & 4173 $\pm$ 64& 2.00 $\pm$ 0.01 & 1.04 $\pm$ 0.15& -0.07 $\pm$ 0.01 &–0.05 $\pm$ 0.39\\  
             $\varepsilon$ For&HD~18907 &K2V& 5044 $\pm$ 103 & 5123 $\pm$ 78& 3.37 $\pm$ 0.24 & 3.52 $\pm$ 0.07& -0.68 $\pm$ 0.08 &–0.60 $\pm$ 0.10\\  
             $\varepsilon$ Eri&HD~22049 &K2Vk& 5130 $\pm$ 69 & 5076 $\pm$ 30& 4.58 $\pm$ 0.09 & 4.60 $\pm$ 0.03& -0.08 $\pm$ 0.05 &–0.09 $\pm$ 0.06\\  
             $\delta$ Eri&HD~23249 &K1III-IV& 5100 $\pm$ 93 & 4954 $\pm$ 26& 3.79 $\pm$ 0.20 & 3.75 $\pm$ 0.02& 0.05 $\pm$ 0.08 &0.06 $\pm$ 0.05\\   
             $\alpha$ CenB&HD~128621 &K1V& 5296 $\pm$ 6 & 5231 $\pm$ 20& 4.59 $\pm$ 0.01 & 4.53 $\pm$ 0.03& 0.26 $\pm$ 0.005&0.22 $\pm$ 0.10\\   \hline
             $\tau$ Cet&HD~10700 &G8.5V& 5366 $\pm$ 109 & 5414 $\pm$ 21& 4.43 $\pm$ 0.18 & 4.49 $\pm$ 0.01& -0.48 $\pm$ 0.08 &–0.49 $\pm$ 0.03\\   
             $\xi$ Hya&HD~100407 &G7III& 5168 $\pm$ 5 & 5044 $\pm$ 38& 3.18 $\pm$ 0.01 & 2.87 $\pm$ 0.02& 0.09 $\pm$ 0.004&0.16 $\pm$ 0.20\\   
             $\mu$ Ara&HD~160691 &G3IV-V& 5856 $\pm$ 7 & 5974 $\pm$ 60& 4.27 $\pm$ 0.01 & 4.30 $\pm$ 0.03& 0.28 $\pm$ 0.004&0.35 $\pm$ 0.13\\   
             Sun&Sun &G2V& 5826 $\pm$ 11 & 5777$\pm$1& 4.42 $\pm$ 0.02 & 4.44 $\pm$ 0.00& -0.02 $\pm$ 0.008&0.0300 $\pm$ 0.05\\   
             $\alpha$ CenA&HD~128620 &G2V& 5914 $\pm$ 5 & 5792 $\pm$ 16& 4.37 $\pm$ 0.01 & 4.30 $\pm$ 0.01& 0.24 $\pm$ 0.004&0.26 $\pm$ 0.08\\   
             18Sco&HD~146233 &G2V& 5891 $\pm$ 12 & 5810 $\pm$ 80& 4.46 $\pm$ 0.02 & 4.44 $\pm$ 0.03& 0.06 $\pm$ 0.007&0.03 $\pm$ 0.03\\   
             -&HD~165499 &G2Va& 5916 $\pm$ 117 & 5676& 4.15 $\pm$ 0.16 & 4.29& -0.12 $\pm$ 0.07 &-0.07\\  
             $\beta$ Hyi&HD~2151 &G0V& 5863 $\pm$ 124 & 5873 $\pm$ 45& 3.91 $\pm$ 0.20 & 3.98 $\pm$ 0.02& -0.11 $\pm$ 0.09 &–0.04 $\pm$ 0.06\\   
             $\beta$ Vir&HD~102870 &G0V& 6203 $\pm$ 11 & 6083$\pm$41& 4.10 $\pm$ 0.01 & 4.10$\pm$0.02& 0.13 $\pm$ 0.007&0.24$\pm$0.07\\  \hline 
             -&HD~45588 &F9V& 6208 $\pm$ 16 & 6168& 4.11 $\pm$ 0.02 & 4.27& -0.05 $\pm$ 0.01 &0.08\\   
             -&HD~156098 &F8IV& 6389 $\pm$ 179 & 6273& 3.69 $\pm$ 0.31 & 3.815494& 0.05 $\pm$ 0.10 &0.075\\  \hline
         \end{tabular} 
         \tablefoot{Final derived $\it{T_{\mathrm{eff}}}$, $\log g$ and [M/H] for the ISAM stars separated into M, K, G and F spectral type. Benchmark LTE parameters from \citet{jofre_gaia_2015} were used for all stars, except for HD~45588 and HD~156098, for which we used values from \cite{van_leeuwen_gaia_2022}. These parameters were used for comparison with ISAM, along with their associated uncertainties (labelled as $_{\rm{ref}}$).}

   \end{table*}

\vspace{-0.5cm}
\section{Measurement of Mg isotopes}\label{sec:3}

\subsection{MgH regions}\label{sec:regions}

In Table \ref{tab:Regions} we have listed 10 MgH features used for Mg isotope fitting from the literature. Regions 1, 2 and 3 were defined by \cite{mcwilliam_isotopic_1985}. Region 4 defined by \cite{melendez_rise_2009}, 5 by \cite{thygesen_chemical_2016}, and the remaining regions by \cite{mckenzie_complex_2024}. We evaluated each region for its reliability for extracting isotope abundances. Regions deemed unreliable were excluded from the analysis, as is discussed in Section \ref{sec:region_removal}. 

\begin{table}[!htb]
      \caption[]{Regions used for magnesium isotopic analysis.}
         \label{tab:Regions}
         \centering
         \begin{tabular}{ccccc}
        
            \noalign{\smallskip}
            Region & Line centre & Lower Wave & Upper Wave & Valid \\
            & \AA & \AA & \AA \\\hline \hline
            \noalign{\smallskip}
            
            \noalign{\smallskip}
        1$^{1}$ & 5134.6 & 5134.42 & 5134.85  & Yes\\
        2$^{1}$ & 5138.7 & 5138.55 & 5138.95  & No \\ 
        3$^{1}$ & 5140.2 & 5140.04 & 5140.46 & Yes\\ 
        4$^{2}$ & 5134.2 & 5134.00 & 5134.40 & No\\
        5$^{3}$ & 5135.1  & 5134.90 & 5134.40 & Yes\\ 
        6$^{4}$ & 5136.1 & 5135.90 & 5135.90 & Yes\\ 
        7$^{4}$ & 5136.4 & 5136.20 & 5136.60  & No\\ 
        8$^{4}$ & 5138.4 & 5138.20 & 5138.60 & Yes\\ 
        9$^{4}$ & 5141.0  & 5141.00 & 5141.45 & Yes\\ 
        10$^{4}$ & 5133.2 & 5133.00 & 5133.40 & Yes\\
            \noalign{\smallskip}
            
         \end{tabular}
         \tablefoot{Regions used for Mg isotopic analysis, labelled by region number for identification. Regions one to three are standard regions for Mg isotopic determination as used in \citet{mcwilliam_isotopic_1988}$^{1}$, while Regions four to ten were introduced in subsequent studies by \citet{melendez_rise_2009}$^{2}$, \citet{thygesen_chemical_2016}$^{3}$, and \citet{mckenzie_complex_2024}$^{4}$. The `Valid' column indicates `Yes' if the region was used in this study and `No' if it was not.}
         \vspace{-0.5cm}
   \end{table}

\vspace{-0.5cm}
\subsection{Isotope pipeline}\label{sec:isotope_pipeline}

While recent isotopic measurements have focused on giant stars and globular clusters \citep{mckenzie_complex_2024}, our study focuses on field giants and dwarfs, expanding the scope of magnesium isotope analysis. To support this, we developed a Python wrapper \footnote{\url{https://github.com/qai11/Isotope-Pipeline}} for the MOOG spectral synthesis code \citep{sneden_moog_2012} that iteratively converges on a best-fit solution, following previous methodologies. Spectral synthesis is central to this process: the pipeline determines magnesium isotopic abundances by performing a $\chi^2$ test between the synthetic and observed spectrum across a range of macroturbulence velocities (V$_{mac}$) and relative abundances of each magnesium isotope. These parameters are varied simultaneously to identify the best fit.

To reduce degeneracy and isolate the isotopic signal, several stellar parameters were fixed during the analysis: effective temperature ($T_{\mathrm{eff}}$), surface gravity ($\log g$), metallicity ([M/H]), projected rotational velocity ($v \sin i$), and $\alpha$-element abundance ([$\alpha$/Fe]) were adopted from the literature \citep{jofre_gaia_2015} for GBS and \citet{van_leeuwen_gaia_2022} for non-GBS stars.

To find the abundances of Mg, europium (Eu) and barium (Ba) we used the standard spectrum fitting method. This process isolates spectral lines from the rest of the spectrum, fitting a synthetic spectra to each observed line using iSpec \citep{blanco-cuaresma_ispec_2015}. Each is checked visually for quality, and an average weighted by the uncertainty of the resulting abundances was calculated (see Section \ref{sec:uncertainties}). 

Global magnesium abundances were determined using eight Mg lines (see Table~\ref{tab:Line_list}) selected from the \textit{Gaia}-ESO survey that had quality flags of Y (yes) or U (uncertain or undecided), indicating suitability for reliable abundance measurements \citep{heiter_atomic_2021}. The derived Mg abundances show good agreement with literature values (see Table \ref{tab:Mg_abunds}), with most stars falling within 2$\sigma$ of the GBS \citep{jofre_gaia_2015}. Note that $\alpha$ Cet and $\tau$ Cet exhibit differences at the 3$\sigma$ level. For $\alpha$ Cet, the discrepancy corresponds to a difference of 0.78 dex, although this remains within 3$\sigma$ due to the relatively large uncertainties in both this work and the GBS reference value. In contrast $\tau$ Cet has a discrepancy of 0.12 dex. These results suggest that the Mg abundances derived here are generally robust.

For some MgH features, minor adjustments to the continuum placement were necessary due to blending with strong Fe and C features in the surrounding spectrum. The Fe abundance was fixed to the literature [M/H] value \citep{jofre_gaia_2015}, and the carbon abundance was determined. Note that CO and C$_2$ abundances were adjusted manually to better fit the wings of some features, improving continuum placement and reducing computation time. This same procedure was applied to Ba and Eu abundances for the neutron-capture element comparison (see Section~\ref{sec:abundances}) using the lines listed in Table~\ref{tab:Line_list}. To maintain consistency across elements, we used the Mg abundances derived in this work rather than relying on literature values, as reference abundances for Ba and Eu were not available for all stars.

Using the global Mg abundance, the individual isotopic abundances were calculated from the isotopic ratios using Equation~\ref{Eqn:Convert_eqn_abund_to_iso_abund}. See Section~\ref{sec:calculating isotopes} for more details. 

 \begin{table}
    \caption{Lines used for abundance determination.}
    \label{tab:Line_list}
     \centering
    \begin{tabular}{cccc} 
     Mg I $\AA$ & C I $\AA$ & Ba II $\AA$ & Eu II $\AA$\\\hline \hline
     5167.322 & 5052.144 & 5853.676 & 6645.0980\\
     5172.684  & 5380.325 & 6141.718 &  -\\
     5183.604  & 6587.610 & 6496.901& -\\
     5528.405 & - & -& - \\
     5711.088 & - & -& - \\
     6318.717 & - & -& -\\
     6319.237 & - & -& -\\
     6319.493 & - & -& -\\
    \end{tabular}
    \tablefoot{Ba, Eu, and Mg lines used for abundances for the stars in the ISAM. Retrieved from \textit{Gaia}-ESO flag list \citep{heiter_atomic_2021}.}
    \vspace{-0.5cm}
\end{table}

With all other parameters fixed, only $V_{\rm mac}$ and the relative abundances of $^{24}$Mg, $^{25}$Mg, and $^{26}$Mg were allowed to vary. Note that $V_{\rm mac}$ accounts for large-scale turbulent motions in a star’s atmosphere that broaden spectral lines beyond the effects of thermal motion or rotation \citep{iliadis_nuclear_2007}. In spectral synthesis and line profile fitting, $V_{\rm mac}$ is a key parameter influencing the shape and depth of absorption features. Rather than fixing $V_{\rm mac}$ to a single value for each star, it was allowed to vary per feature to account for potential differences in how turbulence affects lines of different strengths, excitation potentials, or formation depths. The derived $V_{\rm mac}$ values agreed with typical values for stars of these spectral types \citep{gray_observation_2005}. 

The isotope pipeline operates by exploring the surrounding parameter space from an initial starting point. A synthetic spectrum is generated using these initial values, and its $\chi^2$ is calculated against the observed spectrum. Beginning with $V_{\rm mac}$, we sample nearby values in abundance space and identify the one that minimises the $\chi^2$. This value is then fixed, and the process is repeated sequentially for $^{24}$Mg, $^{25}$Mg, and finally for $^{26}$Mg, yielding a preliminary set of parameters. We continue this iterative sampling of neighbouring values in parameter space until the $\chi^2$ is minimised across all parameters, ensuring convergence on the final solution.

\begin{figure}[!htb]
    \centering
    \includegraphics[width=0.45\textwidth]{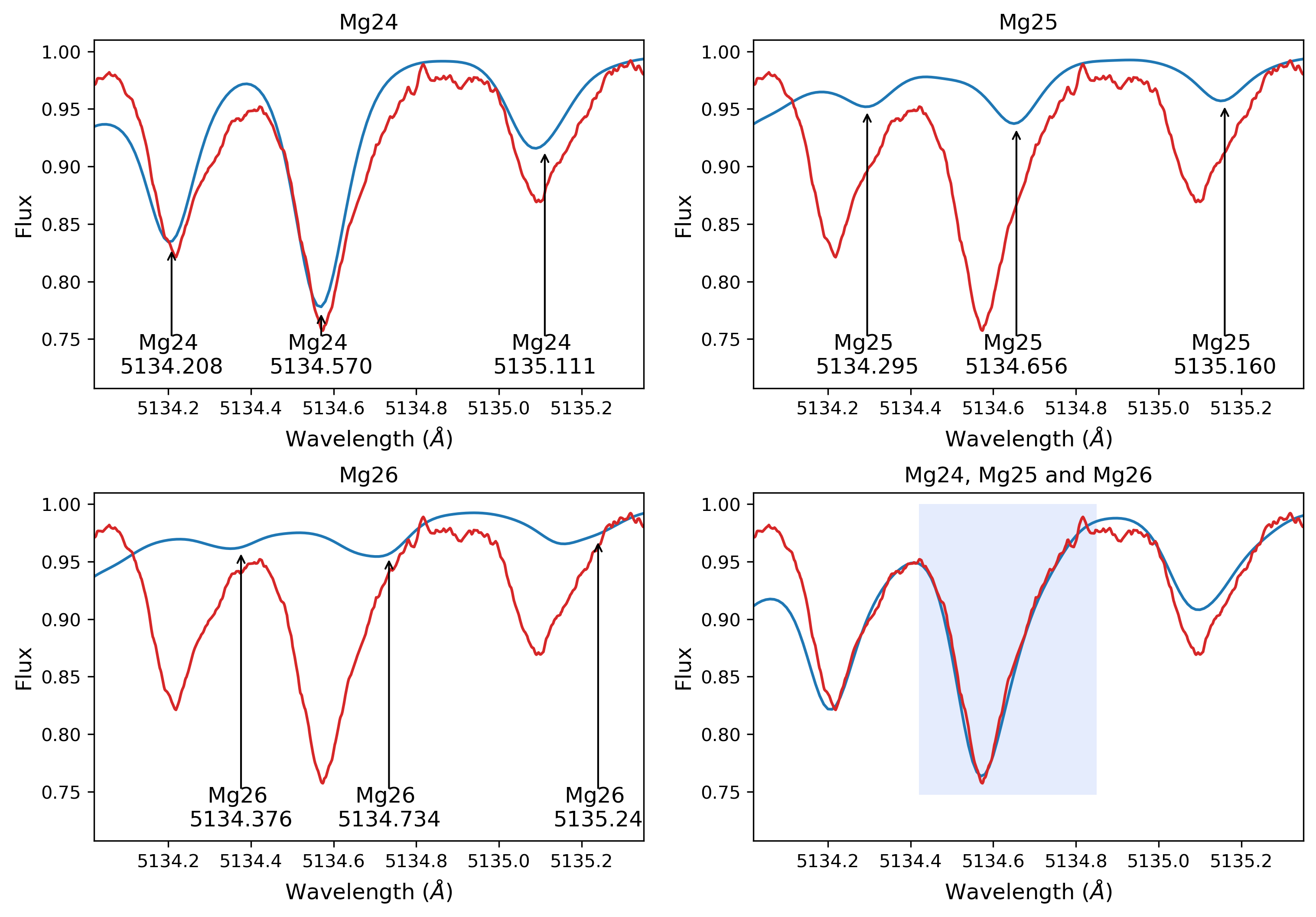}
    \caption{Individual and combined contributions of Mg isotopes to region 1 of HD 18907. The top left panel shows the synthetic spectral feature from $^{24}$Mg only (no contribution from $^{25}$Mg or $^{26}$Mg). The top right panel shows the feature from $^{25}$Mg only, and the bottom left panel shows the $^{26}$Mg contribution. The bottom right panel with the fitting region in the blue box, presents the combined synthetic spectrum including all three isotopes, providing the best fit solution for this region.}
    \label{fig:Isotope_contributions}
\end{figure}

As an example, Figure~\ref{fig:Isotope_contributions} shows the best fit for region 1 of HD~18907. Individual panels illustrate the contribution of each Mg isotope, with the bottom right panel showing the final combined synthetic spectrum. Fits were evaluated only within the coloured region, following the same procedure used for general abundance analysis as described earlier in Section~\ref{sec:isotope_pipeline}. The strongest contribution of the feature comes from $^{24}$Mg, which is why the isotope fitting process begins with $^{24}$Mg after $V_{\rm mac}$.

To minimise potential bias and ensure consistency across the sample, the same initial conditions were used for all stars. These consisted of the approximate solar Mg isotope ratio \citep{de_bievre_table_1985} and the solar $V_{\rm mac}$ of $V_{\rm mac} = 8.41$~km~s$^{-1}$ \citep{mckenzie_magnesium_2024}. The initial isotope ratio was set to $^{24}$Mg:$^{25}$Mg:$^{26}$Mg = 77.69:10.36:11.95 (approximately 2:15:13 in MOOG input), as reported by \citet{mckenzie_magnesium_2024}.

We implement a two-stage fitting approach (coarse and fine) to avoid the solution becoming trapped in a local minimum, which could prevent convergence to the correct isotopic abundances. Using a large initial step allowed the solution to approach the global minimum, and then a finer step refined the isotopic abundances to better represent the observed spectrum. The first iteration used coarse steps (0.5) for the $^{24}$Mg inverse abundance (inverse abundance relative to the isotopic abundance for MOOG), followed by a finer search with steps of 0.1 to refine the fit.

This refinement was applied only to $^{24}$Mg because it is the dominant isotope and contributes most significantly to the MgH molecular features in the spectrum. Due to its stronger influence on the overall line strength and profile, small changes in $^{24}$Mg abundance have a more noticeable effect on the fit quality. In contrast, the contributions from $^{25}$Mg and $^{26}$Mg are minor and primarily affect the shape of the line profile rather than its depth. Once an optimal value for $^{24}$Mg was determined, the sensitivity of the fit to changes in $^{25}$Mg and $^{26}$Mg was reduced, making a coarse search unnecessary for those isotopes. 

Although the isotopic ratios derived from different features within a star did not always agree precisely, they generally clustered around a consistent range. For each iteration, the $\chi^{2}$ values, resulting isotope ratios, and the fit between the observed and synthetic spectra were visually inspected to ensure they made physical sense.

On a per star basis, a spectral region was removed from consideration if any of four conditions were met: the synthetic spectrum did not match the depth of the observed spectrum; the isotope ratios were different by more than 3$\sigma$ with those from other MgH regions in the same star; the $^{24}$Mg ratio differed by more than $25\%$ from other regions; and the $^{25}$Mg and $^{26}$Mg isotopes were not approximately equal (as these isotopes tend to be produced together in similar amounts \citep{iliadis_nuclear_2007}). Additionally, if the core of the molecular feature was well fitted but the wings deviated beyond the fitting region, the line was considered with caution. 

After applying these criteria, a final check of the $\chi^{2}$ was performed to ensure the fit was minimised as much as possible. A difference of up to $25\%$ in the $^{24}$Mg fraction was considered acceptable in order to retain as many regions as possible; however, up to $25\%$ may cause the final isotopic ratio for each star to diverge from the `true’ isotope ratio. Therefore, this condition was applied carefully and only when evaluating a small number of successfully fitted regions. This approach allowed us to derive robust isotopic ratios across the sample, while accounting for blending, continuum placement, and fitting uncertainties.

\vspace{-0.5cm}
\subsubsection{Calculating Mg isotopic abundances}\label{sec:calculating isotopes}

The isotope pipeline determines a ratio for all three isotopes from a predetermined global magnesium abundance. To convert the isotope ratio into individual isotopic abundances, we first calculate the relative ratios of the three Mg isotopes using the inverse ratio approach,
\vspace{-0.76cm}
\begin{center}
    \begin{equation}
    R_k = 
    \frac{\displaystyle \frac{1}{\, i_k}}
    {\displaystyle \sum_{j \in \{24,25,26\}} \frac{1}{\, i_j}}
    \times 100,
    \label{Eqn:isotope_calc}
    \end{equation}
\end{center}

\vspace{-0.2cm}
where $i_k$ represents the measured isotope value for isotope $k$, expressed as an inverse abundance represented as a fraction of the isotope over the total Mg (e.g. $i_{24} = 2$ for $^{24}$Mg). The term $\frac{1}{0.01 , i_k}$ converts the percentage ratio into a scaled inverse value, which reflects the relative contribution of isotope $k$ in the normalisation process. The denominator sums these scaled inverse abundances over all magnesium isotopes ($j \in {24,25,26}$), ensuring proper normalisation. The final result is multiplied by 100 to express $R_k$ as a percentage of the total magnesium isotopic composition. This formulation ensures that the three ratios $R_{24}$, $R_{25}$, and $R_{26}$ sum to 100\%.

Once the isotope ratios are determined, the individual isotopic abundances for each star are calculated using,
\vspace{-0.2cm}
\begin{equation}
    [\textrm{Mg}_i/\textrm{H}] = [\textrm{Mg}/\textrm{H}] + \log_{10}\left( \frac{f_i^{*}}{f_i^{\odot}} \right),
    \label{Eqn:Convert_eqn_abund_to_iso_abund}
\end{equation}

\vspace{-0.1cm}

where $f_i^{*}$ is the fraction of magnesium in isotope $i$ in the star, and $f_i^{\odot}$ is the corresponding fraction in the Sun and [\textrm{Mg}/\textrm{H}] is the global abundance of Mg relative to solar. The same equation is used to propagate the uncertainty.
\vspace{-0.5cm}
\subsubsection{Uncertainty estimation} \label{sec:uncertainties}

Uncertainties in the isotopic abundances were estimated using the Hessian matrix, which captures the gradients of the $\chi^2$ surface around the best-fit solution. The matrix was constructed by evaluating the second derivatives of $V_{\rm mac}$, and the inverse abundances of $^{24}$Mg, $^{25}$Mg, and $^{26}$Mg. To compute these derivatives, each parameter was incrementally shifted up and down, and the resulting change in $\chi^2$ was used to assess the local curvature. Once the Hessian was assembled, its inverse was taken to obtain the covariance matrix, with the square root of the diagonal elements providing the uncertainties.

This method is sensitive to the choice of step size: if the shifts are too large or too small, the matrix can become numerically unstable, leading to inflated uncertainties \citep{cowan_statistical_1998}. The most stable configuration was found to be: $V_{\rm mac} = 0.0001$, $i_{24} = 0.05$, $i_{25} = 0.1$, and $i_{26} = 0.1$. These values produced consistent and stable uncertainty estimates across all fitted regions. 

To convert uncertainties in inverse isotope ratios to actual isotope ratios, we applied the transformation described in Equations~\ref{Eqn:sigma_Pk}–\ref{Eqn:sigma_Rk}, which define the propagation of uncertainty through the inverse ratio formulation.

In addition, uncertainties in the stellar parameters and abundances were computed using iSpec, which applies a non-linear least-squares fitting algorithm (Levenberg–Marquardt) \citep{blanco-cuaresma_determining_2014}. iSpec automatically computes a covariance matrix during fitting, providing a fast and reliable estimate of parameter uncertainties.

The calculation of uncertainties in the magnesium isotope ratios follows from the inverse ratio formulation used in MOOG. These uncertainties are derived from the Hessian matrix and propagated through the ratio transformation.

The inverse ratio term is defined as,

\vspace{-0.5cm}
\begin{equation}
    P_k = \frac{1}{0.01 \cdot i_k}
    \label{Eqn:Pk}
\end{equation}

The uncertainty for each isotope is calculated as,
\vspace{-0.5cm}

\begin{equation}
    \sigma_{P_k} = \left(P_k\right)^2 \cdot \sigma_{i_k}
    \label{Eqn:sigma_Pk},
\end{equation}
where \( i_k \) is the MOOG-derived ratio for isotope \( k \), and \( \sigma_{i_k} \) is its uncertainty from the Hessian matrix of each isotope. 

The uncertainty in the total inverse sum ($\sigma_S$) is then,
\vspace{-0.5cm}

\begin{equation}
    \sigma_S = \sqrt{\sum_k \sigma_{P_k}^2}
    \label{Eqn:sigma_s},
\end{equation}
where $S$ is $\sum_{j \in \{24,25,26\}} \frac{1}{0.01 \, i_j}$.

The normalised isotope ratios are computed as,
\vspace{-0.5cm}

\[
R_k = \left(\frac{P_k}{S}\right) \times 100,
\]
as described in Equation~\ref{Eqn:isotope_calc}.

Finally, the uncertainty in each ratio is propagated using,
\vspace{-0.2cm}
\begin{equation}
    \sigma_{R_k} = R_k \cdot \sqrt{
        \left(\frac{\sigma_{P_k}}{P_k}\right)^2 +
        \left(\frac{\sigma_S}{S}\right)^2
    }
    \label{Eqn:sigma_Rk}
\end{equation}

This formulation ensures that uncertainties in the fitted inverse abundances are accurately reflected in the final isotopic ratios, maintaining consistency with the MOOG fitting framework.

\subsection{MgH lines in FGK spectra}\label{sec:MgH_temps}

\begin{figure}
    \centering
    \includegraphics[width=0.38\textwidth]{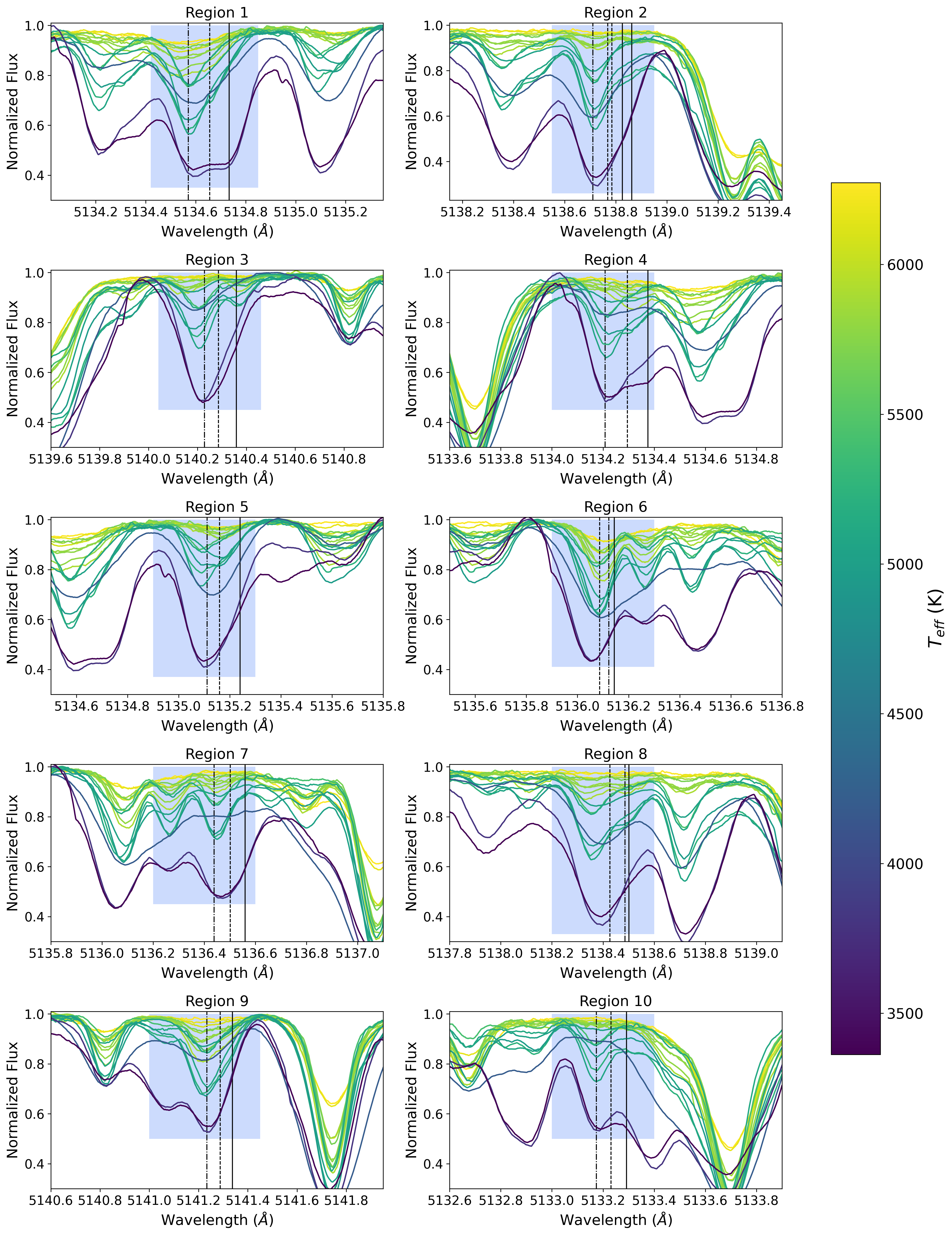}
    \caption{Regions 1 through 10 showing the MgH lines and the fitting region (blue shaded rectangle) for all stars colour coded by effective temperature. Vertical lines mark the positions of the three magnesium isotopes: 
    dot–dash for $^{24}\mathrm{Mg}$, dashed for $^{25}\mathrm{Mg}$, 
    and solid for $^{26}\mathrm{Mg}$.}
    \label{fig:Temp_regions}
\end{figure}

Figure \ref{fig:Temp_regions} shows the 10 MgH regions described in Section \ref{sec:regions}, colour-coded by temperature. It illustrates how the MgH features vary across stars of different temperatures: in the hottest star (lighter colours), HD~156098 ($T_{\mathrm{eff}} = 6273~K$; \citealt{van_leeuwen_gaia_2022}), the features are either very weak or have vanished entirely, whereas in the coolest star (darker colours), HD~11695 ($T_{\mathrm{eff}} = 3362~K$; \citealt{jofre_gaia_2015}), they are much stronger. On the basis of this, a temperature cut-off was identified, above which the isotope pipeline cannot reliably measure magnesium isotopes using the devised methods. This limit is approximately $\sim5400K$, meaning that for nearly all stars hotter than this threshold (primarily G and F type stars) these MgH regions are not viable for determining the Mg isotope abundances using our method. The value of $\approx 5800 \pm 300~K$, that we calculated in Section~\ref{sec:intro} is higher than the observational limit determined in this study. Our dataset includes stars, both above and below this temperature range, for which reliable Mg isotope measurements could not be obtained. This suggests that additional factors may influence detectability, such as $\log~g$, where higher gravity promotes molecular formation, and metallicity or Mg abundance can increase the depth of the spectral features \citep{gray_observation_2005}.

\vspace{-0.5cm}
\section{Analysis and discussion}\label{sec:discussion}

\subsection{Analysis pipeline verification}

This study involved a multi-stage analysis with several iterations and refinements. At each stage, and whenever a new star was added to the sample, we reviewed the pipeline and applied the insights gained, leading to the removal of some regions after assessing their performance across all stars. To assess the reliability of the final version of our isotope analysis pipeline, we compared our results for three stars with published magnesium isotope measurements: $\tau$ Ceti (HD~10700), $\alpha$ Centauri B (HD~128621), and $\delta$ Eridani (HD~23249). Isotopic ratios were determined using a $\chi^2$ minimisation approach across multiple regions (Table~\ref{tab:Regions}), followed by visual inspection of the observed and synthetic spectra. The median weighted averages across all regions were then computed to produce the final isotopic ratios.

\begin{table*}[!htb]
    \centering
\caption{Mg isotope ratios found in this study compared to the literature.}
\label{tab:Reference_isotopes}
\resizebox{\textwidth}{!}{%
    \begin{tabular}{ccccccccc}
           ID1 &ID2&  SPT& This study &B85 & B87 &  ML88 & GL00  &MM24\\ 
           \hline \hline
           $\delta$ Eri  &HD~23249&  K1III-IV& 66:13:22 ($\pm$0.86:0.55:0.58)&- &79:10:11 ($\pm5\%$)&   -&- &-\\
           $\alpha$ Cen B &HD~128621&  K1V &   67:17:16 ($\pm$1.77:0.34:0.5) &-&79:10:11 ($\pm3\%$) &   78:11:11&- &-\\
           $\tau$ Ceti  &HD~10700&  G8.5V&   78:9:12 ($\pm$1.74:2.87:0.5) &84:7:7 ($\pm3\%$)&-&  83:7:10 ($\pm2\%$) & 75:15:9  &-\\
           \hline
 -& HD~55458& K1V& -& -& -& -& 75:15:9&-\\
 -& HD~25329& K1Vb& -& -& -& -& 85:8:8&-\\
 -& HD~149661& K1V& -& -& -& -& 74:12:13&-\\
 -& HD~114095& G8V& -& -& -& -& 79:13:8&-\\
 III-52& -& K3III& -& -& -& -& -&83:13:4\\ 
    \end{tabular}  
    }
    \tablefoot{Comparison of Mg isotope ratios with uncertainties derived in this study with literature values (\citet{barbuy_magnesium_1985}(B85),\citet{barbuy_magnesium_1987}(B87),\citet{mcwilliam_isotopic_1988}(ML88), and \citet{gay_isotopic_2000}(GL00)) for the three reference stars HD~23249, HD~128621, and HD~10700. Additional rows include literature values for stars with similar stellar parameters from \citet{gay_isotopic_2000}(GL00) and \citet{mckenzie_complex_2024}(MM24).}
    
\end{table*}

In Table~\ref{tab:Reference_isotopes}, we compare the magnesium isotope ratios derived in this study with the values reported in the literature for three reference stars, as well as for stars with similar stellar parameters. Our result for $\tau$ Ceti is consistent with the range of literature values with all three isotopes within uncertainty of at least one of the reference values. Deviations are seen in $\alpha$ Centauri B and $\delta$ Eri, with differences of up to 13\% (of total abundance) for $^{24}$Mg and 11\% for $^{26}$Mg. These differences may arise from our use of a larger number of spectral regions, which provides a broader sampling of isotopic features. Additionally, extracting isotopic ratios for the two dwarf reference stars is challenging, as they lie near the upper $T_{\mathrm{eff}}$ limit of the pipeline. Additional literature values were included for comparison, featuring three K-type stars, one G-type star, and one giant star with similar stellar parameters to our targets. The comparison stars are similarly distributed in magnesium isotope parameter space, making them suitable analogues for assessing the consistency and reliability of our measurements.

Overall, agreement with the literature is reasonable (see Table~\ref{tab:Reference_isotopes}), particularly considering that earlier studies relied on manual fitting techniques and smaller number of regions. Previous studies either did not specify which spectral regions were used or focused solely on Regions 1–3 \citep{gay_isotopic_2000}, whereas this work includes Regions 1, 3, 5, 6, 8, 9, and 10. Region 2 and 4 were excluded due to contamination by strong atomic lines in the nearby continuum. The inclusion of additional regions and the use of a $\chi^2$-based fitting procedure likely contributed to the robustness of the derived isotopic ratios. However, direct comparison with the literature is limited by the lack of methodological detail in previous studies, particularly regarding their fitting procedures and continuum treatment. As such, the reliability of our results is best assessed through internal consistency checks and the quality of the fits across multiple regions.

\vspace{-0.5cm}
\subsection{Region quality check}\label{sec:region_removal}

\begin{figure}[!htb]
    \centering
    \includegraphics[width=0.45\textwidth]{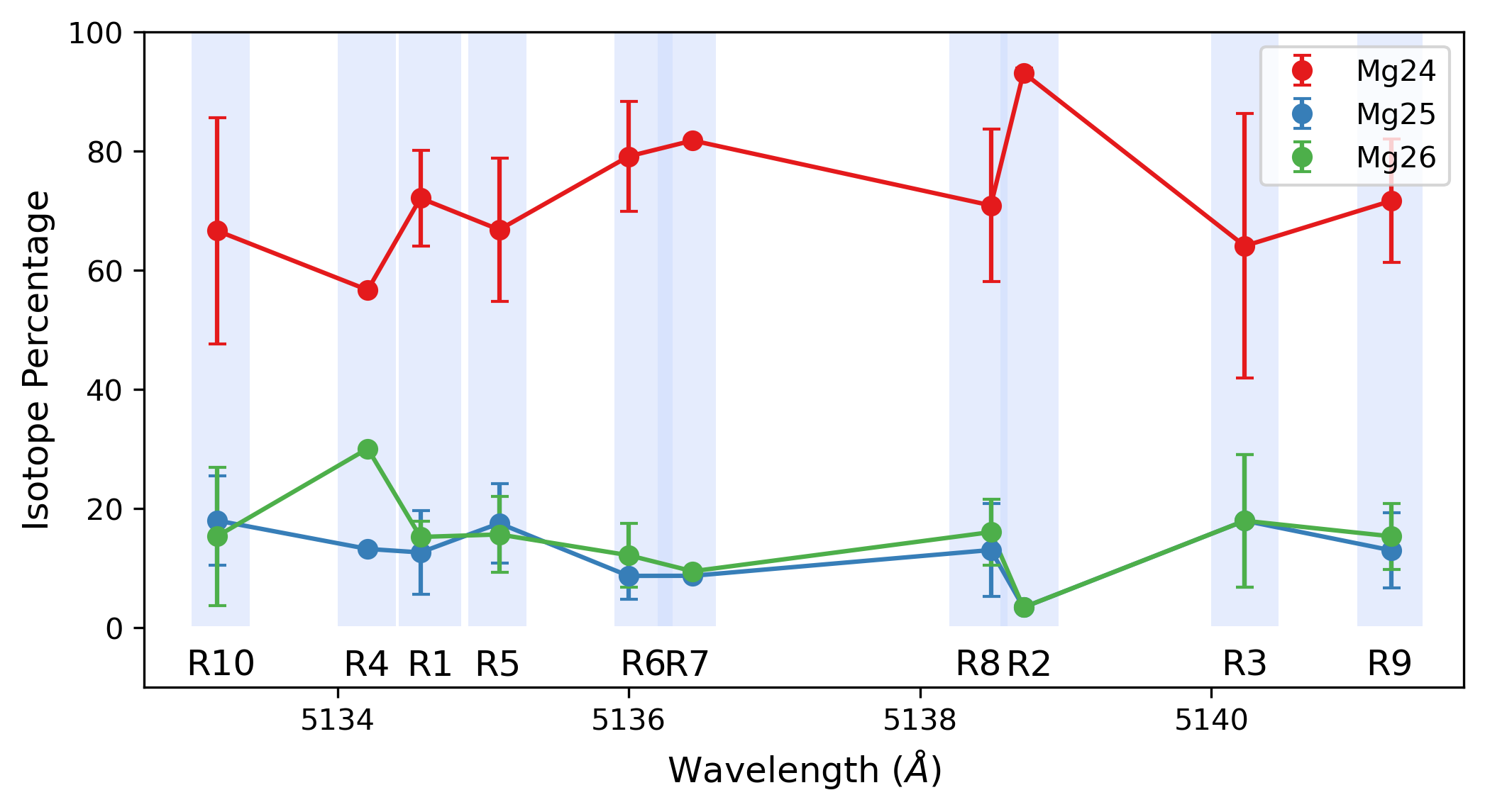}
    \caption{Mean isotopic percentage as a function of region wavelength for $^{24}$Mg (red), $^{25}$Mg (blue), and $^{26}$Mg (green). Error bars indicate the standard deviation of the ISAM for each region.}
    \label{fig:isotope_percentage}
\end{figure}

Figure~\ref{fig:isotope_percentage} shows the isotopic fraction of each magnesium isotope as a function of region wavelength. Each point corresponds to the central wavelength of a region, and the vertical error bars indicate the range of isotopic fraction across the sample. Regions 1, 2 and 3 were used primarily in previous studies \citep{gay_isotopic_2000,yong_magnesium_2003,yong_magnesium_2004}; however, region 2 was excluded in this work due to poor performance. Regions 8 and 10 both have large standard deviations, reflecting significant variation between stars with different magnesium abundances. Regions 1, 5, 6 and 9 show a similar amount of scatter, suggesting they are good for measuring isotope ratios. In contrast, regions 2, 3, 4 and 7 have very little scatter, meaning they are less sensitive to differences in the metallicity, $\log~g$ and $T_{\mathrm{eff}}$ of the stars. It is likely the largest factor in this variation is $T_{\mathrm{eff}}$ as discussed in Section \ref{sec:MgH_temps}. For the ISAM, we recommend using regions 1, 5, 6, 8, 3, 9 and 10 for deriving robust isotopic ratios.

\vspace{-0.5cm}
\subsection{Magnesium isotopes}\label{sec:Mg_abundances}

\begin{table}[!htb]
      \caption[]{Isotope ratios for the ISAM.}
    \label{tab:Isotopes}
         \centering
         \begin{tabular}{ccccccc}
            ID1&$^{24}$Mg&$\delta$$^{24}$Mg&$^{25}$Mg& $\delta$$^{25}$Mg& $^{26}$Mg& $\delta$$^{26}$Mg\\ 
            \hline
            \hline
             $\psi$~Phe & 81.12& 0.06&7.40& 0.02& 11.48& 0.35\\  
             $\alpha$ Cet & 80.84& 0.0062&6.83& 0.0331& 12.32& 0.4472\\  \hline 
             $\varepsilon$ For& 68.53& 0.06&8.52& 0.03& 22.95& 0.38\\  
             $\varepsilon$ Eri&71.12& 0.79& 15.89& 0.30& 12.99& 0.50\\  
             $\delta$ Eri& 65.6& 0.86&12.68& 0.55& 21.72& 0.58\\   
             $\alpha$ CenB&66.91& 1.77&17.15& 0.34& 15.93& 0.50\\   \hline
             $\tau$ Cet&78.42& 1.74&9.57& 2.87& 12.01& 0.50\\   
             $\xi$ Hya&64.50& 0.39&10.02& 2.12& 25.49& 0.50\\
         \end{tabular} 

   \end{table}

\begin{table*}[!htb]
      \caption{Magnesium isotope abundances found in this work.}
    \label{tab:Mg_abunds}
         \centering
         \resizebox{\textwidth}{!}{%
         \begin{tabular}{cccccccccc}
            ID1& [Mg/H]&$\delta$[Mg/H]&ref [Mg/H]&{[$^{24}$Mg/H]}& $\delta$[$^{24}$Mg/H]& [$^{25}$Mg/H]& $\delta$[$^{25}$Mg/H]& [$^{26}$Mg/H]&$\delta$[$^{26}$Mg/H]\\ 
            \hline
            \hline
             $\psi$~Phe & -0.18&0.13& -&-0.17& 0.02& -0.31& 0.07& -0.16&0.15\\  
             $\alpha$ Cet & 0.51&0.24 & -0.270~$\pm$~0.11&0.52& 0.01& 0.35& 0.02& 0.56&0.03\\  \hline 
             $\varepsilon$ For& -0.16&0.10& -0.206~$\pm$~0.05&-0.23& 0.02& -0.23& 0.10& 0.16&0.01\\  
             $\varepsilon$ Eri& 0.01&0.14&-0.078~$\pm$~0.10& -0.04& 0.03& 0.21& 0.11& 0.08&0.29\\  
             $\delta$ Eri& 0.24&0.14& 0.180~$\pm$~0.09&0.16& 0.04& 0.34& 0.11& 0.53&0.13\\   
             $\alpha$ CenB& 0.42&0.15&0.300~$\pm$~0.09&0.35& 0.04& 0.65& 0.04& 0.58&0.24\\\hline
             $\tau$ Cet& -0.09&0.10&-0.224~$\pm$~0.03&-0.10& 0.04& -0.11& 0.12& -0.06&0.18\\ 
             $\xi$ Hya& 0.18&0.09&0.0300~$\pm$~0.11&0.09& 0.09& 0.18& 0.29& 0.55&0.11\end{tabular}
             }
             \tablefoot{Abundances (and uncertainties) of magnesium per star determined from the set of individual Mg spectral lines, and from the isotope pipeline for each magnesium isotope for eight stars calculated using equations in Section \ref{sec:calculating isotopes}. Reference values and uncertainties are from \cite{jofre_gaia_2015} (GBS [Mg/H]). Per isotope abundances are also given.}
            
   \end{table*}

The isotopic ratios determined for the eight stars in this study are presented in Table~\ref{tab:Isotopes}. The results are consistent with various literature sources, as shown in Table~\ref{tab:Reference_isotopes}.

The pipeline failed to produce a consistent isotopic ratio for $\beta$~Ara, a K3Ib-II bright giant \citep{jofre_gaia_2015,iliadis_nuclear_2007}. Both the $\chi^2$ values and visual inspection showed poor agreement with the observed spectrum, likely indicating that $\beta$~Ara lies in a region of parameter space where the pipeline cannot identify any regions that meet the criteria outlined in Section~\ref{sec:isotope_pipeline}. $\beta$~Ara has $[M/H] \approx$ -$0.07$, which is near solar metallicity. This is significantly higher than the metal poor giants seen in \citet{mckenzie_complex_2024} and \citet{thygesen_chemical_2016} which were globular cluster stars of the same spectral type (typically $[M/H] < $–$1$). Increased line blending at near solar metallicity could contribute to the poor fit. Further analysis of a broader sample of bright giants is needed to determine whether this limitation applies across the entire region of parameter space. Therefore, $\beta$~Ara was excluded from further analysis in this study.

The most metal poor star in the sample is $\psi$~Phe ([Fe/H] = -1.24 \citep{jofre_gaia_2015}) and also the coolest ($T_{\mathrm{eff}}\approx3362~K$). The pipeline finds $^{24}$Mg contributing $\sim81\%$ of the total Mg, with excellent agreement showing a standard deviation of 5.5 across regions 2-9. While no literature values are available for this star, the derived ratio is consistent with results from other giants in this study and in the literature \citep{mckenzie_complex_2024}. In Table~\ref{tab:Reference_isotopes} we show III-52 which has a $T_{\mathrm{eff}}\approx4100~K$, $\log g\approx0.51$ and [Fe/H] = -1.7 which has a similar distribution of Mg isotopes.  

This study found that most of the dwarf stars have a slightly lower $^{24}$Mg fraction than the solar value (78.99 \citep{de_bievre_table_1985}), with a mean of 69.9 and standard deviation of 5.3. There is some variation with $^{24}$Mg being slightly lower than other similar spectral type stars in the literature. Variation in $^{24}$Mg is most noticeable for $\alpha$~CenB and $\epsilon$~For (see Tables~\ref{tab:Mg_abunds} and \ref{tab:Reference_isotopes}).

For $\psi$~Phe and $\delta$~Eri, multiple isotopic solutions with similar $\chi^2$ and acceptable visual fits were possible. The pipeline prioritised solutions that were self-consistent across the sample, with $^{24}$Mg contributing most strongly and $^{25,26}$Mg appearing in roughly equal amounts.

\vspace{-0.5cm}
\subsection{Abundances}\label{sec:abundances}

Magnesium isotopes provide information on the chemical history of a star. The $r$ process and $s$~processes occur during the late stages of stellar evolution, ultimately enriching the ISM with heavy elements. The stars observed in this study formed from material that had already been enriched by at least one previous generation of stars. Europium (Eu) and barium (Ba) are commonly used tracers of the $r$ and $s$~processes, respectively \citep{vangioni_cosmic_2019}. To investigate whether magnesium isotopes found in this study reflect the understanding of element nucleosynthesis in the literature, we compared the relative abundances for each Mg isotope with Eu and Ba to look for possible correlations.

Tables~\ref{tab:Eu_Ba_abund} and \ref{tab:Mg_abunds} present the abundances of Eu, Ba, and Mg for the ISAM using the atomic lines in Table~\ref{tab:Line_list}. For Eu and Ba, the reference values available from the Hypatia catalogue \citep{hinkel_stellar_2014} show good agreement (within 1~$\sigma$) with our measurements, despite the relatively large uncertainties in our abundance of Eu. The Mg abundances are all within 2-3 $\sigma$ of the literature values \citep{jofre_gaia_2015} (see Section \ref{sec:isotope_pipeline} for further details).

\begin{table*}
    \caption{Eu and Ba abundances from this work compared to the literature.}
    \label{tab:Eu_Ba_abund}
     \centering
    \begin{tabular}{cccccccccc} 
    ID1& ID2& ref [Eu/H]&  [Eu/H]&  $\delta$ [Eu/H]&  $\Delta$ [Eu/H]&  ref[Ba/H]&  [Ba/H]& $\delta$ [Ba/H]&$\Delta$ [Ba/H]\\ [0.5ex]\hline \hline 
    $\psi$~Phe &HD~11695& - &  -0.29&  0.94&-  & - & 0.59& 0.16&-\\ [0.5ex]
    
    $\alpha$ Cet &HD~18884& - &  0.21&  0.47& - &-  & 0.82& 0.05&-\\ [0.5ex] \hline 

    $\varepsilon$ For&HD~18907&  -0.25~$\pm$~0.1&  -0.35&  1.25&  0.10&  -0.61~$\pm$~0.17& -0.62& 0.05&0.01\\ [0.5ex]
    
    $\varepsilon$ Eri&HD~22049&  0.09~$\pm$~0.16&  0.07&  1.96&  0.02&  0.08~$\pm$~0.18& 0.13& 0.02&0.05\\ [0.5ex]
    
    $\delta$ Eri&HD~23249&  0.23~$\pm$~0.14&  0.20&  0.93&  0.03&  0.06~$\pm$~0.3& -0.01& 0.04&0.07\\ [0.5ex]
    
    $\alpha$ CenB&HD~128621&  0.16~$\pm$~0.18&  0.19&  0.20& 0.03 & 0.16~$\pm$~0.17  & 0.36& 0.13& 0.20\\[0.5ex] \hline
    
    $\tau$ Cet&HD~10700&  -0.02~$\pm$~0.18&  -0.07&  1.77&  0.05&  -0.53~$\pm$~0.18& -0.54& 0.05&0.01\\ [0.5ex]
    
    $\xi$ Hya&HD~100407&  0.13~$\pm$~0.03&  0.28&  0.06&  0.15&  0.23~$\pm$~0.11& 0.50& 0.04&0.27\\ \hline 
    \end{tabular}
    \tablefoot{Eu and Ba abundances for the stars in ISAM measured in this study (including uncertainties, $\delta$), alongside reference values (ref) extracted from the Hypatia catalogue \citep{hinkel_stellar_2014} and the citations therein. The difference ($\Delta$) between measured and reference values is also provided. All abundances are reported using the solar normalisation from \cite{grevesse_solar_2007}, and each reference value represents the median of all entries for that star in the catalogue.}
\end{table*}

\vspace{-0.5cm}
\subsection{$r$ and $s$~process comparison}

The GBS are good reference stars, but they are not necessarily representative of a stellar population. In the following, we explore correlations between magnesium isotope abundances and neutron capture elements. This is typically done with a larger sample in a defined stellar population. 

The $^{24}$Mg isoptope is produced primarily during the $\alpha$ process, with a small contribution from the $r$~process \citep{iliadis_nuclear_2007, vangioni_cosmic_2019}, which occurs in high-mass stars. Therefore, we expect to see some correlation with europium, which is produced predominantly in the $r$~process following the $\alpha$ process \citep{guiglion_ambre_2018,sneden_line_2014}. The $^{25,26}$Mg isotopes are mainly produced in hot-bottom burning that precedes the $s$~process, with minor contributions from the $r$~process \citep{iliadis_nuclear_2007, vangioni_cosmic_2019}. We therefore expect some correlation with barium, which is mostly produced in the $s$~process \citep{guiglion_ambre_2018,sneden_line_2014}, and a much weaker correlation with europium.

To find if there is any correlation between elemental abundance and isotopic abundance, comparisons between each were made using a Pearson \textit{r} correlation coefficient algorithm. The calculated \textit{p}-value was used to confirm or refute any visual trends. The Pearson \textit{r} test uses the null hypothesis, which is the default assumption that there is no effect or correlation between two samples. The \textit{p}-value describes the likelihood of a correlation between, in this case, two data sets. A small \textit{p}-value ($\leq 0.05$) indicates that the observed correlation would be rare if the null hypothesis of no correlation were true, providing evidence for a real association. In contrast, a large \textit{p}-value ($>0.05$) indicates weak evidence against the null hypothesis, meaning that the observed results could plausibly arise by random chance \citep{cowan_statistical_1998}.

\begin{figure*}[!htb]
    \centering
        \centering
        \includegraphics[width=0.7\textwidth]{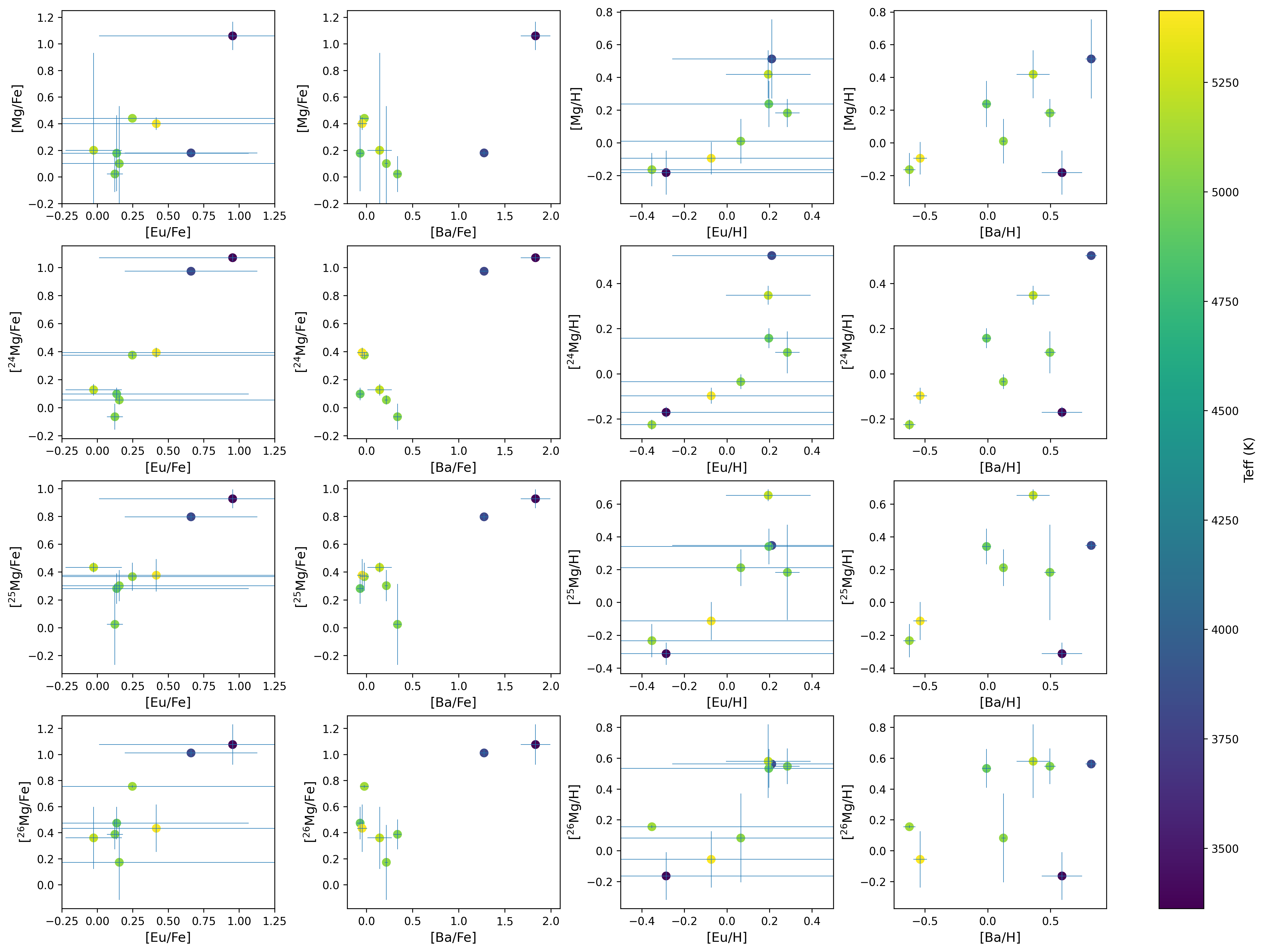}
    \caption{A set of 16 plots showing comparisons of abundances of europium and barium with the isotopic magnesium abundance coloured by $T_{\mathrm{eff}}$. The left eight panels show [${^{24}\mathrm{Mg}}$/Fe], [${^{25}\mathrm{Mg}}$/Fe], [${^{26}\mathrm{Mg}}$/Fe] and the right eight panels show [${^{24}\mathrm{Mg}}$/H], [${^{25}\mathrm{Mg}}$/H] and [${^{26}\mathrm{Mg}}$/H].}
    \label{fig:Eu&Ba vs Mg}
\end{figure*}

In Figure~\ref{fig:Eu&Ba vs Mg} we show the relationships between the `pure' $r$ and $s$~process elements with the three isotopes of Mg. The left six panels show comparisons with [X/Fe] and the right show comparisons with [X/H].

The comparison of [X/H] with the three magnesium isotopes and europium show a strong correlation ($r = 0.795$–$0.86$) that were statistically significant ($p < 0.05$). The strong correlation between $^{24}$Mg and Eu ($r=0.842, p=0.0087$) is consistent with theoretical expectations, as $^{24}$Mg is predominantly produced by hydrostatic $\alpha$ capture in massive stars, a process that precedes the $r$~process responsible for Eu production. There appears to be an upper limit in the [Eu/H]–Mg trend, where [Eu/H] stops increasing around $\sim0.2$ despite continued increase in global Mg and $^{24}$Mg. This plateau is less evident for $^{25}$Mg and $^{26}$Mg. $^{25}$Mg and $^{26}$Mg also show strong positive correlations with Eu ($r=0.86$, $p=0.0061$; $r=0.813$, $p=0.0141$). These isotopes can be produced both by the $r$~process and by the AGB \citep{karakas_mg_2003,vangioni_cosmic_2019}. However, the uncertainties associated with $^{25}$Mg and $^{26}$Mg measurements are substantially larger than those for $^{24}$Mg, which reduces confidence in the observed correlations.

In contrast to the strong correlations observed with [Eu/H], we find little evidence of correlation between [Mg/H] and [Ba/H]. A moderate correlation can be seen between $^{24}$Mg and Ba ($r=0.629, p=0.095$). This is not statistically significant and is unexpected, as Ba and $^{24}$Mg are predominantly synthesised in different stellar environments \citep{karakas_mg_2003,vangioni_cosmic_2019}. However, we see no correlation between Ba and $^{25}$Mg and $^{26}$Mg, despite expectations that all three are predominantly produced in similar astrophysical environments. This discrepancy may be due to the small sample size, differences in the specific subprocesses within the $s$~process (strong and weak $s$~process \citep{karakas_mg_2003,vangioni_cosmic_2019}), or the larger measurement uncertainties associated with the weaker and more blended spectral features of $^{25}$Mg and $^{26}$Mg. If this result reflects a genuine astrophysical trend, it may suggest that only a small amount of $^{25}$Mg and $^{26}$Mg is produced during the strong $s$~process. 

In Figure \ref{fig:Eu&Ba vs Mg} we found that the coolest star in the sample $\psi$~Phe, may be an outlier with the most uncertain parameters and so finding good abundances are very difficult. In this case we removed it from the sample to assess how the correlations differ for [X/H]. In every case we find significant increases in the significance levels for all three isotopes with the largest increase being with $^{25}$Mg. For all isotopes we find a strong correlation $p>0.75$ and $r<0.045$. With removal of this outlier we are seeing what we initially expected when comparing to Ba for the heavy isotopes, $^{25}$Mg and $^{26}$Mg, which is a strong correlation at high significance. This indicates that our results support current theories of nucleosynthetic production sites. For Eu we find that the correlation decreases, where $^{26}$Mg is now weakly significant, however $^{24}$Mg and $^{25}$Mg are still statistically significant.

Direct comparison of [Eu/H] and [Ba/H] against the global [Mg/H] abundance shows a significant correlation. For Mg vs Eu there is a strong positive correlation ($r=0.84, p=0.009$). This suggests that the total Mg is enhanced when Eu is enhanced. Broadly speaking this makes sense because they are both enriched by massive stars and released during Type II Supernovae. Specifically, the largest contributing isotope $^{24}$Mg is produced during hydrostatic $\alpha$-capture in stellar interiors, in a process preceding explosive $r$~process events which Eu is synthesised. For Ba vs. Mg, there is no statistically significant correlation ($r = 0.586$, $p = 0.127$). This lack of correlation is expected, as Ba and Mg are primarily produced in different astrophysical environments \citep{karakas_mg_2003,vangioni_cosmic_2019}. However, some scatter in the data may arise due to galactic chemical evolution (GCE).

Making a comparison to [X/H] may also include effects due to metallicity. To remove any such, the comparisons were repeated with [X/Fe] (see second set of panels in Figure \ref{fig:Eu&Ba vs Mg}). There is some change in how Ba, Eu and Mg interact. A moderate correlation is seen for Mg and Ba ($r=0.618, p=0.103$) and strong significance for Mg and Eu ($r=0.772, p=0.025$). For [Eu/Fe] vs [$^{24}$Mg/Fe] a strong and statistically significant correlation is seen. For [Eu/Fe] vs [$^{25}$Mg/Fe] and [$^{26}$Mg/Fe] we see the same strong correlations ($r=0.849, p=0.007$; $r=0.848, p=0.007$). A second outlier is prominent which weakens the presence of any correlation between Eu and $^{25}$Mg. For the comparisons of [Ba/Fe] to Mg and its isotopes, strong correlations are found which are dominated by the two cooler stars. Removal of these data points gives a strong negative correlation in all cases. We can conclude that the [X/Fe] show there is an effect due to metallicity but the final conclusion remains the same, we need more data to be convinced of any real correlations.

\vspace{-0.5cm}
\subsection{Abundance vs parameters}

\begin{figure}
    \centering
        \centering
        \includegraphics[width=0.45\textwidth]{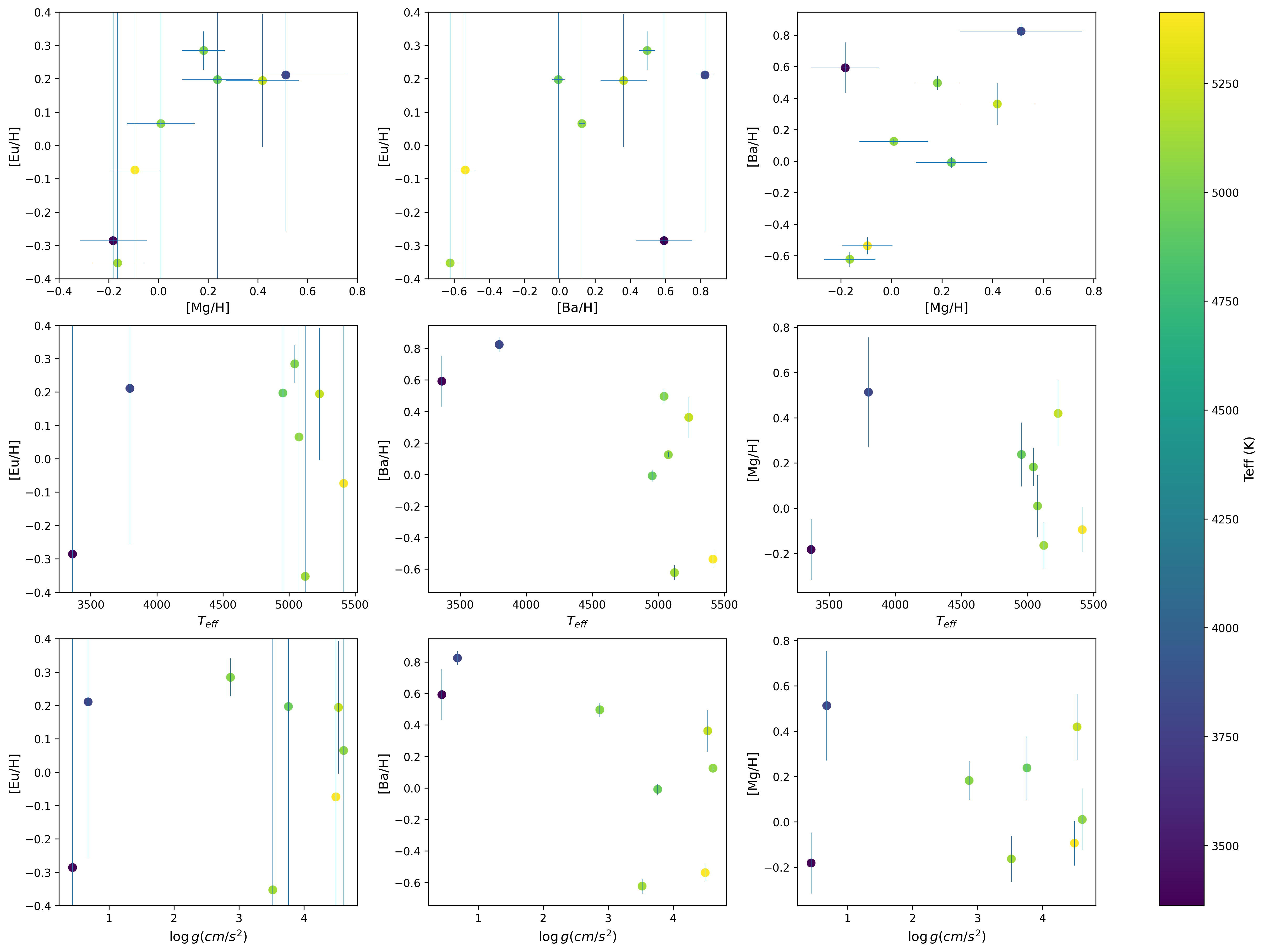}
    \caption{Abundances of europium, barium, and magnesium compared with stellar parameters to examine possible trends coloured by $T_{\mathrm{eff}}$.}
    \label{fig:Eu&Ba&Mg vs params}
\end{figure}

In Figure~\ref{fig:Eu&Ba&Mg vs params} we show the relationship between Eu, Ba and Mg against the stellar parameters of $T_{\mathrm{eff}}$ and $\log g$.When comparing the neutron capture elements of Ba and Eu to the stellar parameters there is little correlation ($r=0.50, p=0.20$). The same can be said with the Mg to the stellar parameters ($T_{\mathrm{eff}}$:$r=-0.03, p=0.93$, $\log g$:$r=-0.08, p=0.86$). There is a near significant trend for Ba vs $T_{\mathrm{eff}}$ and Ba vs $\log g$. For Ba vs $T_{\mathrm{eff}}$ a $r=-0.67$ and $p=0.066$ show a fairly convincing negative correlation with a near significant \textit{p}-value. This indicates that there may be a correlation but more data is needed to confirm the result. For Ba vs $\log g$ a $r=-0.65$ and $p=0.081$ there is also a fairly strong negative correlation with a near significant \textit{p}-value. These both indicate that higher $T_{\mathrm{eff}}$ and lower surface gravity might show more Ba but it is possible that this is an observational or systematic effect. There is no significant correlation between the Ba and Eu abundances which is expected since they are generated in different environments \citep{vangioni_cosmic_2019}. There is no correlation between Mg and $T_{\mathrm{eff}}$ or $\log~g$, suggesting that these abundances are robust to stellar parameter bias. With removal of metallicity bias the same robust results are seen.

\vspace{-0.5cm}
\section{Conclusion}\label{sec:conclusion}

We have developed an analysis pipeline for determining magnesium isotopic ratios in stars across a range of spectral types. Validation with reference stars demonstrates good agreement with literature values, confirming the reliability of the method within its effective temperature limits. Our analysis shows that $^{24}$Mg consistently dominates the isotopic composition, with $^{25}$Mg and $^{26}$Mg contributing roughly equal fractions. Examination of individual MgH regions identified Regions 1, 3, 5, 6, 8, 9, and 10 as providing the most reliable isotopic information, while other regions are less sensitive due to limited variation across stars. Elemental abundances of Mg, Eu and Ba are consistent with previous studies, supporting the accuracy of isotopic determinations \citep{jofre_gaia_2015, hinkel_stellar_2014}. 

Comparison of magnesium isotopes with the $r$~process element Eu and the $s$~process element Ba \citep{guiglion_ambre_2018,sneden_line_2014} reveals strong correlations with [Eu/H] but weak or absent correlations with [Ba/H]. A comparison removing outliers resulted in stronger correlations for [Ba/H] with all isotopes. The removal of the outlier resulted in findings that are more consistent with theoretical expectations, as the majority of $^{24}$Mg is released by the same process as Eu and $^{25}$Mg, and $^{26}$Mg are produced through processes releasing $s$~process elements \citep{vangioni_cosmic_2019,karakas_mg_2003}. With removal of the effect of metallicity similar results were found. The absence of strong dependence on stellar parameters further supports the robustness of the derived abundances and isotopic ratios. Overall, these results validate the reliability of the analysis pipeline and provide a foundation for future studies of magnesium isotopes across diverse stellar populations, contributing to our understanding of chemical evolution in the Galaxy.

\begin{acknowledgements}
    I would like to thank Kyle Boucher, Peter Cottrell, Michael Elston, Heather Sinclair-Wentworth, and Zachary Lane, for helping me check my arguments. 
        
    This research has made us of the Python 3 packages, \texttt{numpy} \citep{harris_array_2020}, \texttt{pandas} \citep{team_pandas-devpandas_2020}, and \texttt{matplotlib} \citep{hunter_matplotlib_2007}.
\end{acknowledgements}

\vspace{-0.85cm}

\bibliographystyle{aa} 
\bibliography{Bibliography.bib}

\end{document}